\PassOptionsToPackage{bookmarks=false}{hyperref}
\documentclass[sigconf,authorversion]{acmart}

\usepackage{booktabs}
\usepackage{enumitem}
\usepackage{multirow}
\usepackage{footnote}
\usepackage{array}
\usepackage{amsmath}
\usepackage{listings}
\usepackage{color}
\usepackage{colortbl}
\usepackage{balance}
\usepackage{rotating}
\usepackage{tablefootnote}
\usepackage{ifthen}
\usepackage{tikz}
\usepackage{xspace}
\usepackage{soul}
\usepackage{float}

\setitemize{leftmargin=*}

\newif\ifIsDoubleBlind
\IsDoubleBlindfalse

\newif\ifIsIncludeAppendix
\IsIncludeAppendixfalse

\setcopyright{acmlicensed}

\acmDOI{10.1145/3319008.3319021}

\acmISBN{978-1-4503-7145-2/19/04}

\acmYear{2019}
\copyrightyear{2019}

\acmPrice{15.00}

\acmConference[EASE '19]{Evaluation and Assessment in Software Engineering}{April 15--17, 2019}{Copenhagen, Denmark}
\acmBooktitle{Evaluation and Assessment in Software Engineering (EASE '19), April 15--17, 2019, Copenhagen, Denmark}

\input glyphtounicode.tex
\definecolor{ColorWeakLightBlue}{RGB}{220, 240, 255}
\definecolor{ColorHighlightLightBlue}{RGB}{191, 225, 255}
\definecolor{ColorLightGray}{RGB}{220, 220, 220}

\newcolumntype{L}[1]{>{\raggedright\let\newline\\\arraybackslash\hspace{0pt}}m{#1}}
\newcolumntype{R}[1]{>{\raggedleft\let\newline\\\arraybackslash\hspace{0pt}}m{#1}}
\newcommand{\halfBaselinespace}{\vspace{0.6\baselineskip}}
\newcommand{\minispace}{\vspace{0.3\baselineskip}}

\newcommand{\codeText}[1]{\begin{small}\texttt{#1}\end{small}}
\newcommand{\codeTextInFootnote}[1]{\begin{footnotesize}\texttt{#1}\end{footnotesize}}

\newcommand{\gitHash}[1]{\begin{footnotesize}\texttt{#1}\end{footnotesize}}

\newcommand{\studyObject}[1]{\textsc{#1}}
\newcommand{\formatCaptionDetails}[1]{\textmd{\small{#1}}}
\newcommand{\formatResultBox}[1]{\halfBaselinespace \noindent \fbox{\parbox{\linewidth}{#1}} \minispace}

\newcommand{\coloredSquare}[1]{\tikz\filldraw[#1,fill=#1] (0,0) rectangle ++(4pt,4pt);}
\newcommand{\coloredLine}[1]{\textcolor{#1}{\rule[0.5ex]{0.32cm}{1.2pt}}}

\newcommand{\blueText}[1]{\textcolor[rgb]{0,0,1}{#1}}

\begin{document}

\newcommand{\paperTitle}{Is the \mbox{Stack Distance} Between \mbox{Test Case} and Method \mbox{Correlated With} Test Effectiveness?}
\newcommand{\paperTitleShort}{Is Stack Distance Correlated With Test Effectiveness?}

\newcommand{\loc}{LOC\xspace}
\newcommand{\kloc}{kLOC\xspace}
\newcommand{\sdist}{minimal stack distance\xspace}
\newcommand{\sdistShort}{stack distance\xspace}
\newcommand{\sdistPlural}{minimal stack distances\xspace}
\newcommand{\sdistShortCapitalized}{Stack Distance\xspace}
\newcommand{\sdistCapitalized}{Minimal Stack Distance\xspace}

\newcommand{\effectT}{effectively tested\xspace}
\newcommand{\ineffShort}{ineffectively\xspace}
\newcommand{\ineff}{\ineffShort tested\xspace}
\newcommand{\ineffMeth}{\ineff methods\xspace}
\newcommand{\mtRes}{mutation testing result\xspace}
\newcommand{\mtReses}{mutation testing results\xspace}
\newcommand{\predGoal}{the \mtRes of a method\xspace}
\newcommand{\purpose}{identifying \ineffMeth} 
\newcommand{\usage}{an alternative to mutation testing or as a preceding, less costly step to that\xspace}
\newcommand{\usageLong}{a light-weight alternative to mutation testing or as a preceding, less costly step to that\xspace}

\newcommand{\numberOfStudyObjects}{21\xspace}
\newcommand{\numberOfStudyObjectsText}{21\xspace}
\newcommand{\locOfAllStudyObjects}{1,824.9\,k\xspace}
\renewcommand{\locOfAllStudyObjects}{1.8~million\xspace}
\newcommand{\maxTestsNum}{5,254\xspace}
\newcommand{\maxTestsPrj}{\studyObject{Commons Math}\xspace}
\newcommand{\maxLocNum}{240.6\,k\xspace}
\newcommand{\maxLocPrj}{\studyObject{biojava}\xspace}
\newcommand{\minLineCoverageNum}{28.0\%\xspace}
\newcommand{\maxLineCoverageNum}{95.0\%\xspace}
\newcommand{\minBranchCoverageNum}{19.3\%\xspace}
\newcommand{\maxBranchCoverageNum}{91.1\%\xspace}
\newcommand{\minIneffTested}{1.7\%\xspace}
\newcommand{\maxIneffTested}{32.0\%\xspace}

\newcommand{\pit}{Pitest\xspace}
\newcommand{\pitAlternativeName}{PIT\xspace}
\newcommand{\pitVersion}{1.4.0\xspace}
\newcommand{\pitDescartesVersion}{1.2.4\xspace}
\newcommand{\pitMpVersion}{1.3.4\xspace}
\newcommand{\miningAlgorithm}{Random Forest\xspace}
\newcommand{\correlationMethod}{Spearman\xspace}

\newcommand{\sdistMaxCorr}{0.58\xspace}
\newcommand{\predictionWithinMxoForIneffMedianPrecision}{70.7\%\xspace}
\newcommand{\predictionWithinMxoForIneffMedianRecall}{34.3\%\xspace}
\newcommand{\predictionWithinMxoForIneffBestCasePrecision}{96.6\%\xspace}
\newcommand{\predictionWithinMxoForIneffBestCaseRecall}{100.0\%\xspace}
\newcommand{\predictionWithinMxoForBothMedianPrecision}{92.9\%\xspace}
\newcommand{\predictionWithinMxoForBothMedianRecall}{93.4\%\xspace}
\newcommand{\predictionWithinMtoPairForIneffMedianPrecision}{82.4\%\xspace}
\newcommand{\predictionWithinMtoPairForIneffMedianRecall}{71.7\%\xspace}
\newcommand{\predictionWithinMtoPairForBothMedianPrecision}{84.8\%\xspace}
\newcommand{\predictionWithinMtoPairForBothMedianRecall}{85.3\%\xspace}
\newcommand{\predictionCrossMxoForIneffWithSmoteMedianPrecision}{19.2\%\xspace}
\newcommand{\predictionCrossMxoForIneffWithSmoteMedianRecall}{43.2\%\xspace}
\newcommand{\predictionCrossMxoForBothMedianPrecision}{85.6\%\xspace}
\newcommand{\predictionCrossMxoForBothMedianRecall}{88.1\%\xspace}
\newcommand{\slowDownAdaBoost}{about eleven times\xspace}

\ifIsDoubleBlind
\else
	\newcommand{\citeResults}{\cite{Niedermayr2018SDistData}}
	\newcommand{\citeTool}{\cite{Niedermayr2018TestAnalyzer}}
	
	\newcommand{\acknowledgmentText}{
		This work was partially funded by the German Federal Ministry of Education and Research (BMBF), grant ``SOFIE, 01IS18012A''.
		The responsibility for this article lies with the authors.
	}
\fi

\title[\paperTitleShort]{\paperTitle}

\ifIsDoubleBlind
\else
	\newcommand{\institutionCqse}{CQSE GmbH\xspace}
\newcommand{\institutionUniStuttgart}{University of Stuttgart\xspace}

\newcommand{\authorFullNameRN}{Rainer Niedermayr\xspace}
\newcommand{\authorFullNameSW}{Stefan Wagner\xspace}

\newcommand{\authorShortNameRN}{R. Niedermayr\xspace}
\newcommand{\authorShortNameSW}{S. Wagner\xspace}

\newcommand{\authorCityRN}{Garching b. M\"unchen\xspace}
\newcommand{\authorCitySW}{Stuttgart\xspace}

\newcommand{\authorStateRN}{Germany\xspace}
\newcommand{\authorStateSW}{Germany\xspace}

\newcommand{\authorCityStateRN}{\authorCityRN, \authorStateRN\xspace}
\newcommand{\authorCityStateSW}{\authorCitySW, \authorStateSW\xspace}

\newcommand{\authorEmailRN}{niedermayr@cqse.eu}
\newcommand{\authorEmailSW}{stefan.wagner@iste.uni-stuttgart.de}

\newcommand{\authorInstitutionRN}{\institutionUniStuttgart, \institutionCqse\xspace}
\newcommand{\authorInstitutionSW}{\institutionUniStuttgart\xspace}

\newcommand{\authorOrcidIDRN}{0000-0002-9563-6743\xspace}
\newcommand{\authorOrcidIDSW}{0000-0002-5256-8429\xspace}

\author{\authorFullNameRN}
\affiliation{
  \institution{\authorInstitutionRN}
  \city{\authorCityRN} 
  \state{\authorStateRN} 
}
\email{\authorEmailRN}

\author{\authorFullNameSW}
\affiliation{
  \institution{\authorInstitutionSW}
  \city{\authorCitySW}
  \state{\authorStateSW} 
}
\email{\authorEmailSW}

\renewcommand{\shortauthors}{\authorShortNameRN and \authorShortNameSW}

\fi

\begin{abstract}
	Mutation testing is a means to assess the effectiveness of a test suite
 and its outcome is considered more meaningful than code coverage metrics.
However, despite several optimizations, mutation testing requires a significant computational effort
 and has not been widely adopted in industry.
Therefore, we study in this paper whether test effectiveness can be approximated
 using a more light-weight approach.
We hypothesize that a test case is more likely to detect faults in methods
 that are close to the test case on the call stack
 than in methods that the test case accesses indirectly through many other methods.
Based on this hypothesis, we propose the \sdist between test case and method as a new test measure,
 which expresses how close any test case comes to a given method,
 and study its correlation with test effectiveness.
We conducted an empirical study with \numberOfStudyObjectsText open-source projects,
 which comprise in total \locOfAllStudyObjects~\loc,
 and show that a correlation exists between \sdistShort and test effectiveness.
The correlation reaches a strength up to \sdistMaxCorr.
We further show that a classifier using the \sdist along with additional easily computable measures
 can predict \predGoal
 with \predictionWithinMxoForBothMedianPrecision precision and \predictionWithinMxoForBothMedianRecall recall.
Hence, such a classifier can be taken into consideration as \usageLong.

\end{abstract}



\maketitle

\section{Introduction}
\label{Sec:Introduction}

Automated software tests are an important means for quality assurance in software projects and
 are used to reveal faults and prevent regressions in software applications.
Different measures to evaluate test suites have been proposed.
Most common are code coverage metrics~\cite{huang1975approach,zhu1997software}
 expressing which portion of the application code is executed by test cases.
They can be computed at different levels, 
 for example, as line coverage, branch coverage, or decision coverage~\cite{chilenski1994applicability}.
However, since code coverage metrics measure test completeness and do not assess oracle quality,
 they are not necessarily suitable for expressing the test effectiveness of a test suite~\cite{inozemtseva2014coverage,niedermayr2016teststellme,antinyan2018mythical}.
More advanced approaches take data-flow criteria into account~\cite{rapps1982data}
 and measure which portion of the covered statements is checked in assertions~\cite{schuler2013checked}.

Another established, powerful technique to evaluate test suites is mutation testing~\cite{jia2011analysis}.
The general idea behind mutation testing is to generate mutants by seeding faults into the code of a program
 and check whether the tests can kill (detect) these faults.
Hence, compared to code coverage metrics, this technique takes oracle quality into account
 and can provide more meaningful results.
However, mutation testing is---despite several optimization techniques---computationally complex
 due to the effort needed for generating and testing a large number of mutants. 
Despite its effectiveness, there are no indications that mutation testing is widely adopted as a test efficacy criterion in practice~\cite{ivankovic2018industrial, jia2011analysis}.

Since mutation testing can be expensive and code coverage is not necessarily meaningful enough for assessing test suites,
 we study in this paper whether test effectiveness can be approximated using a more light-weight approach.
We hypothesize that a test case that directly invokes a method is more likely to detect faults in that method
 than another test case that accesses the method indirectly through many others.
Therefore, we propose a measure called \textit{\sdist}, which expresses how close any test case comes to a given method,
 and study whether methods with a high \sdist value are more likely to be ineffectively tested.
For that, we conduct a mutation analysis using the Descartes operator~\cite{vera2018descartes}
 and assess whether methods that contain surviving mutants
 exhibit a higher \sdist than the remaining methods.
Furthermore, we train a classifier using stack distance values and further measures,
 which can be collected in a single execution of a test suite,
 and evaluate the classifier's performance in predicting \mtReses.

\textit{Research goal:}
We aim at reducing the effort to identify ineffectively tested code.
In this paper, we investigate how well the \sdistShort measure correlates with and can be used to predict test effectiveness.
This would allow us to use it as alternative to mutation testing.

\textit{Contributions:}
This paper makes two contributions:
First, we propose and study the \sdist measure, which characterizes the proximity of a method to any of its test cases.
Second, we evaluate a machine-learning classifier based on method test-case characteristics
 and show that classifiers to predict \mtReses
 can come into question as \usage.
For example, this could allow the use in continuous integration where mutation testing would take too long
 or is not applicable for other reasons.

The remainder of the present paper is organized as follows.
Section~\ref{Sec:Related_Work} discusses related work.
Section~\ref{Sec:Definitions} defines relevant terms.
Afterwards, Section~\ref{Sec:SDistApproach} describes the approach
 to compute the \sdistShort measure.
Section~\ref{Sec:Study} presents design and results of the empirical study.
Then, Section~\ref{Sec:Discussion} discusses the study's results and implications,
 and Section~\ref{Sec:Threats_To_Validity} explains threats to validity.
Finally, Section~\ref{Sec:Conclusion} summarizes the main findings and sketches future work.

Data to replicate the study is available at~\citeResults.

\section{Related Work}
\label{Sec:Related_Work}

Mutation testing was first proposed by Lipton~\cite{lipton1971fault} in the 1970s
 and formalized by DeMillo et al.~\cite{demillo1978hints}.
It has since then been extensively studied~\cite{jia2011analysis,usaola2010mutation,papadakis2017mutation}.
In general, mutation testing is computationally complex;
 to address this downside, researchers have suggested several approaches to reducing the cost of mutation analysis.
Offutt et al.~\cite{offutt2001mutation} classified these approaches as
 \textit{do fewer,} \textit{do smarter,} and \textit{do faster}.
\textit{Do fewer} approaches comprise
 the use of a smaller, representative set of mutation operators~\cite{offutt1996experimental,offutt1993experimental,siami2008sufficient},
 sampling of mutants~\cite{acree1980mutation},
 mutants clustering~\cite{ji2009novel},
 and higher order mutation, in which multiple mutation operators are applied at once~\cite{jia2008constructing}.
The most prominent \textit{do smarter} approach is weak mutation,
 in which a mutant is immediately evaluated after its execution point instead of checking it at the end of a test execution~\cite{howden1982weak,jia2011analysis}.
\textit{Do faster} approaches comprise further run-time optimization techniques to speed up the generation and execution of mutants
 (e.g., bytecode mutants~\cite{ma2005mujava,schuler2009efficient}, aspect-oriented mutation~\cite{bogacki2006evaluation}, or parallel mutation testing~\cite{fleyshgakker1994efficient}).

In our work,
 we study whether measures describing the relationship between methods and test cases can uncover
 ineffectively tested methods representing surviving mutants.
Hence, we propose an approach to predict \predGoal in a light-weight way without the need for executing mutation testing.

Namin et al.~\cite{siami2008sufficient} used linear models to predict the overall mutation score,
and Jalbert et al.~\cite{jalbert2012predicting} also predicted that score using machine learning models.
However, both did not perform predictions on individual methods.
Strug et al.~\cite{strug2012machine, strug2018evaluation} calculated the structural similarity of mutants,
 predicted based on results of similar mutations whether a given test would detect a mutant or not,
 and thereby reduced the number of mutants to be executed.
However, their approach still requires a mutation analysis of a subset of mutants.
The most related work to ours is from Zhang et al.~\cite{zhang2018predictive},
 who predicted the \mtRes of individual mutations and achieved promising results.
They also included mutations that are not covered by any test case and hence cannot be killed.
In contrast to the work of Zhang et al.,
 we predict \predGoal and not of single mutations,
 exclude methods that cannot be killed since they are not covered,
 and include the proposed \sdist measure in the prediction model.

Stack distance as a measure was first defined and used by Mattson et al. to evaluate storage hierarchies~\cite{mattson1970evaluation}.
Ca\c{s}caval et al. used it to estimate cache misses~\cite{cabetacaval2003estimating}.
Barford et al. used it for web servers to measure the likelihood that a requested file will be requested again in the near future~\cite{barford1998generating}.
In this paper, we define \sdistShort in the context of testing to characterize the proximity between test cases and methods.

\section{Definitions}
\label{Sec:Definitions}

We define the \textit{\sdist between a method $m$ and a test case $t$}
 as the length of the shortest path from $t$ to $m$.\footnote{In
 this paper, we define and apply \sdist based on \textit{methods.}
 However, the definitions are also applicable to \textit{functions} in non-object-oriented programming languages.
}
Hence, the value is one for a method that is directly invoked by a given test case
 and, for example, two for a method that is indirectly invoked by a given test case through one other method.

\begin{figure}
	\centering
	\includegraphics[width=0.8\linewidth]{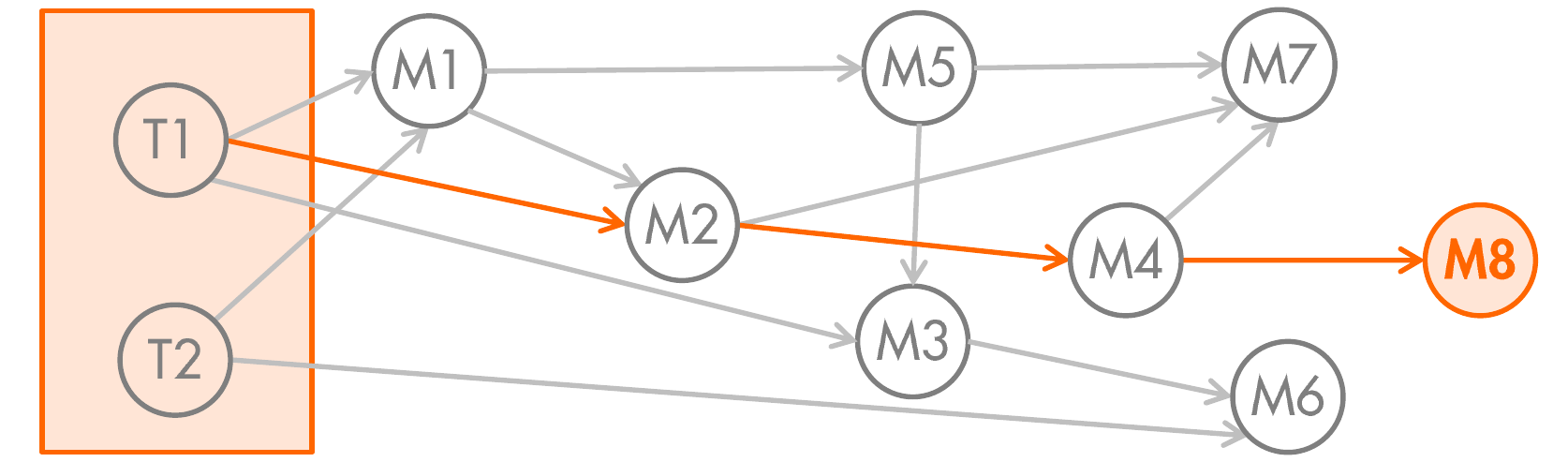}
	\caption{The \sdist of method \textit{M8} is $3$.
	 No test case can access \textit{M8} through fewer method invocations.}
	\label{Fig:SDist_Example}
\end{figure}

We define the \textit{\sdist of a method $m$} as the shortest distance
 between $m$ and any of its covering test cases $T(m)$.
It corresponds to the minimal distance on the call stack
 between the method $m$ and all test cases.
Figure~\ref{Fig:SDist_Example} illustrates an example.

We call a method \textit{covered} if it is executed by at least one test case.
The \textit{\mtRes} of a covered method can either take the value \textit{\ineff} or \textit{\effectT}.
We consider a covered, non-empty method as \textit{\ineff}
 if its whole logic can be removed without causing any test case to fail.
Such \ineffMeth are also known as \textit{pseudo-tested} methods~\cite{niedermayr2016teststellme,vera2018descartes}.
The idea behind pseudo-testedness is that if no single test case can detect such an extreme transformation,
 test cases will not be able to detect more subtle mutations.
\label{Sec:Descartes_Operator}
Pseudo-tested methods can be detected with the Descartes mutation operator, which works as follows~\cite{vera2018descartes}.
For void methods, the operator removes the whole method body.
For methods with a return type, depending on the type, one or two mutants are created,
 which replace the method body with a statement returning a value satisfying the declared return type.
Table~\ref{Tbl:Descartes_Return_Values} presents the return values per type.
When two mutants are created, a method is only considered pseudo-tested if both mutants cannot be killed;
 hence, the use of two mutants avoids that equivalent mutants influence the \mtRes of a method.
\begin{table}
	\centering
	\caption{Return values of the Descartes operator.}
\begin{tabular}{lll}
\toprule
Return Type Class       & Mutant 1          & Mutant 2 \\
\midrule
 void										& \textit{(void)}			& \textit{(not created)} \\
 boolean 								& \codeText{false} 		& \codeText{true} \\
 byte, short, int, long & \codeText{0} 				& \codeText{1} \\
 float, double 					& \codeText{0.0} 			& \codeText{0.1} \\
 char 									& \codeText{' '} 			& \codeText{'A'} \\
 string 								& \codeText{""} 	  	& \codeText{"A"} \\
 T[]		 								& \codeText{new T[]\{\}} 	  	& \textit{(not created)} \\
 reference type					& \codeText{null} 	  	& \textit{(not created)} \\
\bottomrule
\end{tabular}
	\label{Tbl:Descartes_Return_Values}
\end{table}

We further use common mutation testing terms as defined in literature~\cite{jia2011analysis}:
A \textit{mutation operator} is a transformation rule
 that generates a \textit{mutant} by applying syntactical changes to the original program.
A mutant is said to be \textit{killed} if at least one test case of the test suite fails due to the changes;
 otherwise it is said to have \textit{survived}.
An \textit{equivalent mutant} is---despite syntactical changes---semantically equivalent to the original program
 and can therefore not be killed.

\section{Computation of \sdistCapitalized}
\label{Sec:SDistApproach}

\begin{figure}
	\centering
	\includegraphics[width=1.0\linewidth]{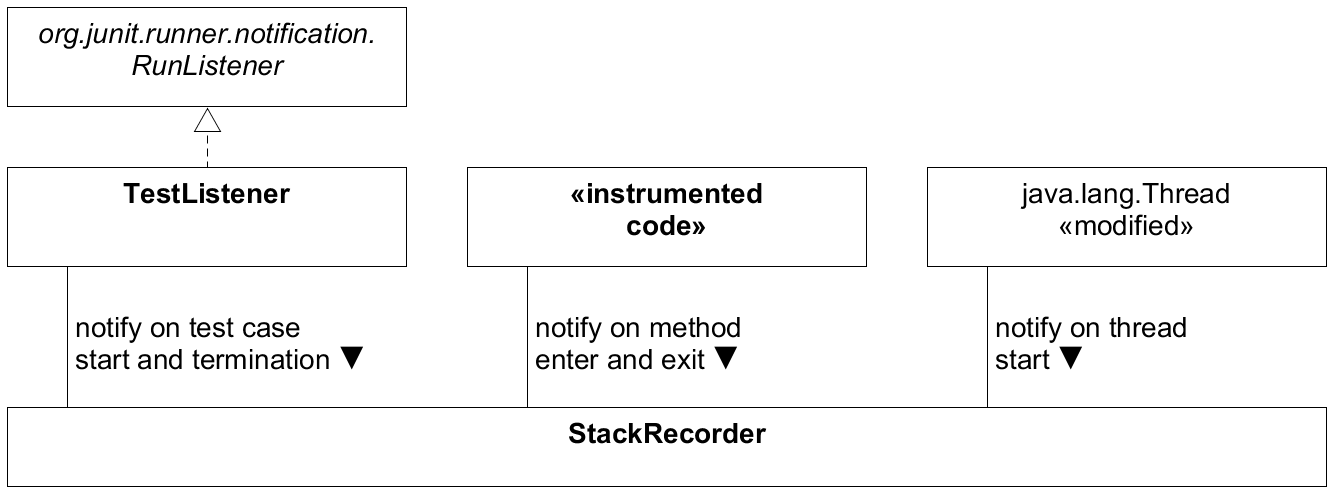}
	\caption{Overview of the \sdistShort computation.}
	\label{Fig:SDist_Recording}
\end{figure}

In the following, we describe the computation of the \sdist for Java applications;
 nonetheless, this measure is applicable to other programming languages as well.
The steps to compute the \sdist comprise the instrumentation of the code,
 the replacement of Java's \codeText{Thread} class,
 and the recording of the method invocations during the test execution.
Figure~\ref{Fig:SDist_Recording} presents an overview of the computation.

\textit{1) Instrumentation:}
We instrument each method of the source code so that it notifies our stack-recorder class
 when a method is entered and exited.
To instrument a method, we introduce a new try-finally block
 and move the original code into the try block.
We then insert a statement before the try block,
 which calls our recorder class with the signature of the considered method.
Next, we insert a further statement into the newly created finally block,
 which informs the recorder that the method invocation needs to be removed from the current stack.
The finally block is always invoked when the method is left (even if an exception is raised or propagated).

To conduct the code instrumentation, we developed a Maven-plugin,
 which operates at the byte-code level and uses the ASM\footnote{\url{http://asm.ow2.io/}} library.
The decompiled source code of an instrumented method might look as follows:
\begin{lstlisting}
  public int getSize() {
    InvocationLogger.push(
         "org.SampleClass.getSize()");
    try {
      /* BEGIN ORIGINAL CODE */
      return this.size;
      /* END ORIGINAL CODE */
    } finally {
      InvocationLogger.pop(
         "org.SampleClass.getSize()");
    }
  }
\end{lstlisting}

\textit{2) Thread class replacement:}
To achieve a thread-aware computation of the \sdist,
 we need to be aware of the current stack height of each thread
 and know which thread was started by which other thread.
For that, we need to be notified when a new thread is started.
Since Java's \codeText{Thread} class does not provide the possibility to register listeners,
 we took the original code from the JDK and adjusted it
 so that our stack-recorder class gets informed about the start of a new thread.
We compiled the modified thread class and put it into the ``endorsed'' folder of the JDK. 
The replacement of the thread class does not influence test results.

\textit{3) Recording:}
Finally, we need to execute the test suite and record the distances between test cases and methods.
We use Maven's Surefire plugin for the execution of unit tests and Failsafe plugin for integration tests
 and register our stack-recorder class as test listener in these plugins.
Hence, the recorder will be notified when a new test case execution begins
 and can assign all subsequent method invocations to that test case.
When a test case execution starts and an instrumented method is entered,
 the method's signature is pushed onto the recorder's stack for the current thread.
Then, the stack's height is counted and, if appropriate, the distance from the executed test case to the start of the current thread is added.
If the resulting distance constitutes a new minimum for a given method test-case pair,
 the pair's \sdist value is updated.
When an instrumented method is left,
 its signature is taken down from the stack of the appropriate thread.

Note that if a method is invoked recursively, the height of the stack increases with each invocation;
 however, we are only interested in the \textit{minimal} stack distance of each method test-case pair.

In short, the recorder class holds the so far \sdist of each executed method test-case pair,
 the method invocations on the stack of each thread,
 and the relations between the threads.
At the end of each test case execution, the \sdist values are persisted.

Note that another imaginable approach
 that computes the stack height by requesting the current thread to dump its stack trace (as done when creating exceptions)
 is not fast enough to be viable for doing the computation in test executions.

Limitations are as follows:
We applied the instrumentation to all methods except constructors.
We excluded constructors,
 because it is tricky to instrument a constructor in a way so that its beginning is correctly intercepted,
 because a constructor's very first statement unavoidably delegates to another constructor or a super constructor
 such that the code there gets executed first.
Consequently, constructor invocations will not be counted when computing the \sdistShort;
 notwithstanding the above, methods invoked by constructors are still considered.
Furthermore, external libraries are not instrumented;
 therefore, method invocations in external libraries are not counted. 
The consequence of both limitations is that the computed \sdistShort will in some cases be slightly lower than the actual distance.
Hence, the computed \sdistPlural should be considered as a lower bound.

\section{Empirical Study}
\label{Sec:Study}

This section reports on the design and results of the empirical study
 that we conducted to investigate the influence of the \sdist between test case and method on test effectiveness.
We further examined how well the \mtRes of a method can be predicted using this measure.

\subsection{Research Questions}

\newcommand{\rqNumSDist}{RQ\,1\xspace}
\newcommand{\rqNumPrediction}{RQ\,2\xspace}

\newcommand{\rqTextSDist}{Are methods with a higher stack distance to the test cases more likely to be \ineff?\xspace}
\newcommand{\rqTextPrediction}{How well can \predGoal be predicted using test-relationship measures?\xspace}

We investigate the following research questions:

\halfBaselinespace

\textbf{\rqNumSDist: \rqTextSDist}
With this research question, we want to find out
 whether the \sdist of a method is correlated with the property how well a method is tested.
We hypothesize that a test case that never comes close to a given method
 is not effective in detecting faults in that method.
Consequently, we expect a method tested only by distant test cases to be less effectively tested.
In other words, we hypothesize that methods with a high \sdist are more likely to contain surviving mutants.
The answer to this question helps determining whether \sdistShort can be a useful predictor for test effectiveness.

\halfBaselinespace

\textbf{\rqNumPrediction: \rqTextPrediction}
Since mutation testing is costly,
 we want to find out whether a more light-weight approach can approximate results gained from mutation analysis.
We are interested in predicting \predGoal based on measures characterizing relationships between methods and test cases.
If such a prediction approach works well, it could be used as \usage.

\subsection{Study Objects}

\begin{table*}
	\centering
	\caption{Study objects.}
	\begin{tabular}{llR{0.9cm}R{0.9cm}R{1.55cm}R{1.55cm}R{1.55cm}}
		\toprule
		 Name $ \downarrow $														&	Purpose									&	\loc  & \#\,Tests		& Line Cov.	& Branch Cov. & Git Revision \\
		\midrule

\studyObject{Apache Commons Geometry}	&	geometric utilities	&	19.4\,k	&	643	&	76.9\%	&	70.7\%	&	\gitHash{be34ad93}	\\
\studyObject{Apache Commons Imaging}	&	image library	&	48.4\,k	&	575	&	71.3\%	&	58.9\%	&	\gitHash{eb98398b}	\\
\studyObject{Apache Commons Lang}	&	utility classes for Java	&	77.0\,k	&	4,053	&	95.0\%	&	91.1\%	&	\gitHash{1f0dfc31}	\\
\studyObject{Apache Commons Math}	&	mathematics library	&	186.3\,k	&	5,254	&	89.8\%	&	84.8\%	&	\gitHash{eafb16c7}	\\
\studyObject{Apache Commons Statistics}	&	statistics library	&	6.1\,k	&	358	&	91.5\%	&	87.6\%	&	\gitHash{aa5cbad1}	\\
\studyObject{biojava}	&	biological data processing	&	240.6\,k	&	1,181	&	40.5\%	&	38.5\%	&	\gitHash{523c78e1}	\\
\studyObject{bitcoinj}	&	Java Bitcoin library	&	59.1\,k	&	5,222	&	67.5\%	&	61.3\%	&	\gitHash{911f6d49}	\\
\studyObject{geometry-api-java}	&	spatial data processing	&	87.0\,k	&	408	&	71.6\%	&	59.4\%	&	\gitHash{3704c220}	\\
\studyObject{google-gson}	&	JSON serialization	&	14.8\,k	&	1,039	&	84.4\%	&	79.2\%	&	\gitHash{57085d62}	\\
\studyObject{Google HTTP Java Client}	&	HTTP client library	&	30.1\,k	&	635	&	54.9\%	&	58.8\%	&	\gitHash{df0e9f2a}	\\
\studyObject{graphhopper}	&	route planning library and server	&	60.5\,k	&	1,680	&	65.4\%	&	60.9\%	&	\gitHash{e954f008}	\\
\studyObject{jackson-databind}	&	databinding for JSON data	&	103.0\,k	&	2,159	&	77.8\%	&	70.7\%	&	\gitHash{bf604125}	\\
\studyObject{javaparser}	&	parser and AST for Java	&	118.4\,k	&	1,284	&	59.8\%	&	48.1\%	&	\gitHash{1cca4c46}	\\
\studyObject{JFreechart}	&	chart library	&	222.8\,k	&	2,175	&	55.5\%	&	46.4\%	&	\gitHash{39dfee3c}	\\
\studyObject{jsoup}	&	HTML and CSS parser	&	18.2\,k	&	671	&	81.4\%	&	77.8\%	&	\gitHash{220b7714}	\\
\studyObject{openwayback}	&	web wayback machine	&	66.8\,k	&	320	&	28.0\%	&	26.8\%	&	\gitHash{680fba15}	\\
\studyObject{pdfbox}	&	PDF document manipulation	&	227.6\,k	&	1,587	&	49.7\%	&	43.3\%	&	\gitHash{d9930344}	\\
\studyObject{scifio}	&	scientific image format IO	&	79.4\,k	&	1,019	&	37.1\%	&	19.3\%	&	\gitHash{281e7ce2}	\\
\studyObject{traccar}	&	server for GPS tracking	&	59.6\,k	&	310	&	56.4\%	&	49.0\%	&	\gitHash{6d259427}	\\
\studyObject{urban-airship}	&	library for marketing platform	&	37.9\,k	&	706	&	79.3\%	&	46.0\%	&	\gitHash{98edb3ca}	\\
\studyObject{vectorz}	&	fast vector mathematics	&	61.9\,k	&	456	&	61.1\%	&	63.8\%	&	\gitHash{a05c69d8}	\\

		\bottomrule
	\end{tabular}
	\label{Tbl:Study_Objects}
\end{table*}

We selected study objects from GitHub\footnote{\url{https://github.com}} based on the following criteria:
The projects need to be written in Java,
 contain test cases designed for the JUnit test framework,
 and use Maven as build system.
We manually selected five Apache projects (\studyObject{Commons Geometry}, \studyObject{Commons Imaging}, \studyObject{Commons Lang}, \studyObject{Commons Math}, \studyObject{Commons Statistics}), and \studyObject{JFreechart},
 which are popular open-source projects used in several empirical test studies
 (e.g., in \cite{hemmati2015effective,inozemtseva2014coverage,just2014defects4j}).
We selected additional study objects that satisfy the previously mentioned criteria
 by searching GitHub for recently updated projects with at least five forks (to require a certain popularity).
We excluded a project
 if it was not possible to build it (e.g., due to compilation problems or unresolvable dependencies),
 if more than 5\% of the test cases failed in a local execution of the original test suite,
 or if the mutation analysis was not successful (e.g., due to special test runners or class loading mechanisms).

The selected study objects are from different domains and contain both single- and multi-module projects.
Their characteristics are presented in Table~\ref{Tbl:Study_Objects}.
\textit{\loc} (lines of code) refers to the application code (i.e., code without test and sample code)
 and was measured with Teamscale~\cite{heinemann2014teamscale}.
\textit{\#\,Tests} refers to the number of test cases as reported by Maven.
\textit{Line} and \textit{branch coverage} were computed with JaCoCo\footnote{\url{https://www.eclemma.org/jacoco/}}.
The largest project, \maxLocPrj, consists of \maxLocNum~\loc.
\maxTestsPrj contains with \maxTestsNum the most test cases.
The line coverage of the projects ranges between \minLineCoverageNum and \maxLineCoverageNum.

\subsection{Study Design}
\label{Sec:Study_Design}

\textbf{\rqNumSDist:}
We hypothesize that the higher the \sdist of a method is to any test case,
 the less likely the method is effectively tested.
To test this hypothesis, we analyze whether a correlation exists
 between a method's \sdist to any test case and its \mtRes (i.e., whether a method is \ineff by \textit{all} test cases or not).
For that, we compute for each project the \correlationMethod rank correlation coefficient,
 which expresses the strength of this relationship (between $-1$ and $+1$), and the p-value.
We use a significance level of 0.05.
Moreover, we present plots illustrating the proportion of \ineffMeth per \sdist value.

\halfBaselinespace

\textbf{\rqNumPrediction:}
To answer this research question, in which we train and evaluate a classifier to predict \mtReses,
 we collect further measures besides \sdistShort for each covered method.
We chose the following method measures because they 
 can easily be computed during a single execution of a test suite:

\begin{itemize}[topsep=0pt]
  \item Line count: number of coverable lines of code in the method
	\item Branch count: number of branches
	\item Line coverage: proportion of covered lines out of coverable lines
	\item Branch coverage: proportion of covered branches out of coverable branches (100\% for covered methods without branches)
	\item Number of covering test cases: number of test cases that execute the method
	\item Scope of covering test cases: minimum number of covered methods of any of the method's covering test cases
	\item Maximum invocation count: maximum number of invocations of the method during the execution of any covering test case
	\item Return type of the method: void, boolean, numeric, string, array, reference to object
\end{itemize}

\halfBaselinespace

For each project, we train one machine-learning classifier
 to predict \predGoal with respect to all covering test cases,
 and one to predict \predGoal test-case pair.

We evaluate the performance of the models with respect to
 within-project and cross-project predictions.
Within-project evaluations show how well predictions work when models are trained on a data-subset of the same project,
 cross-project evaluations indicate how well models can be generalized to conduct predictions in other projects.
For within-project predictions, we apply repeated ten-fold cross-validation~\cite{kohavi1995study}.
For cross-project predictions, we test each project with a model that is trained on the respective remaining projects. 

We measure model performance by computing precision, recall, and F-score.
Following Zhang et al.~\cite{zhang2018predictive}, we predict both outcomes (\ineffShort and \effectT)
 and use the weighted average of the performance metrics
 (i.e., ``each metric is weighted according to the number of instances with the particular class label'').
In addition, we report the performance of the outcome \textit{\ineff},
 because methods with this outcome represent the minority class and are therefore more difficult to predict.

Furthermore, we exemplary show the prediction model's computed variable importances for one project.

\subsection{Data Collection and Processing}
To collect data for the study,
 we first executed the test suite of each study object and recorded the \sdist of each method test-case pair.
The recording of the \sdistShort was carried out as defined in Section~\ref{Sec:Definitions}
 and described in Section~\ref{Sec:SDistApproach}.
Note that we were working on the existing test suites of the projects; we did not generate test cases.

\begin{table}
	\centering
	\caption{Example of a full mutation matrix.}
\begin{tabular}{ccr}
\toprule
Method       & Test Case          & Mutation Testing Result \\
\midrule
$m_1$	& $t_1$ & \ineff \\
$m_1$	& $t_2$ & \effectT \\
$m_2$	& $t_2$ & \ineff \\
\bottomrule
\end{tabular}
	\label{Tbl:Full_Mutation_Matrix}
\end{table}

Second, we conducted a mutation analysis for each study object.
For that, we used \textit{\pit} (\pitAlternativeName)~\cite{coles2016pit} in version \pitVersion
 with the \textit{pit-mp} extension to support multi-module projects.
\pit is a well-known mutation testing tool for Java applications
 and has been used in several studies (e.g., \cite{ahmed2016testedness,gopinath2015hard,gopinath2016limits}).
As performance optimization, \pit aborts the analysis of a mutant after the mutant is first killed by a test case.
However, for this study, we need a full mutation matrix,
 which contains the result (killed or survived) of each mutant for each covering test case.
Therefore, we adjusted \pit to compute a full mutation matrix as proposed by~\cite{ahmed2016testedness}.
Table~\ref{Tbl:Full_Mutation_Matrix} presents an example of such a matrix.

To gain further insights, we made an additional adjustment to \pit
 and recorded for each killed mutation by what event it was killed.
Hence, we know for a mutation whether it is detected by a test case
 because of a failing assertion (\codeText{AssertionError})
 or because of another implicit exception being thrown
 (e.g., \codeText{Null\-Pointer\-Exception}, or \codeText{ArithmeticException} due to a division by zero).

We used \pit with the \textit{Descartes} plugin~\cite{vera2018descartes},
 which implements the mutation operator to uncover pseudo-tested methods
 (see Section~\ref{Sec:Descartes_Operator}).
More details on the mutation operator can be found in~\cite{niedermayr2016teststellme}.

We excluded empty methods and methods solely returning \codeText{null} from the analysis because their mutation would result in an equivalent mutant.
We also excluded \codeText{hashCode} methods
 because we are convinced that mutation testing is not suitable for assessing their testing state.\footnote{As
  long as a \codeTextInFootnote{hashCode} method considers no additional fields for computing an object's hash value,
  it still fulfills its contract even if fewer fields are considered or another computation formula is used.
  The used mutation operator does not introduce additional field accesses.
 }
We further excluded constructors because, as described in the limitations of the \sdistShort computation in Section~\ref{Sec:SDistApproach},
 we cannot compute reliable \sdistShort values of these special methods.
Moreover, we excluded generated code, which was present for example in \studyObject{bitcoinj},
 because the code is re-generated during the build process and not designed to be tested.

For \rqNumPrediction, we collected further measures to enhance the prediction model.
We used \textit{JaCoCo} to compute a method's number of lines and branches as well as line and branch coverage values.
The number of covering test cases per method and their scope was computed based on the full mutation matrix.
The method's invocation count during a test execution was collected alongside the \sdistShort recording.
Finally, the return type of a method was deduced from the mutation testing output.

We used the statistical software \textit{R} to process data.
We trained and evaluated prediction models with R's \textit{caret} package~\cite{caret2017doc}.
We chose \textit{\miningAlgorithm} as machine-learning algorithm
 because preliminary experiments on our datasets revealed that it achieved the best performance.
\textit{adaboost} achieved an almost equal performance, but was \slowDownAdaBoost slower.
Zhang et al. also used \miningAlgorithm for their predictions~\cite{zhang2018predictive}.

\subsection{Results}

This section presents the results to the research questions.
Data to reproduce the results are available at \citeResults.

\begin{table}
	\centering
	\caption{Overview of the mutation analysis results.}
	\begin{tabular}{lR{1.95cm}R{3.6cm}}
		\toprule
		Project						& \# \ineffMeth & \% \ineff \mbox{out of all} \textit{covered} methods $ \downarrow $ \\
		\midrule

\studyObject{scifio}	&	154	&	32.0\%	\\
\studyObject{pdfbox}	&	829	&	26.3\%	\\
\studyObject{biojava}	&	1147	&	24.4\%	\\
\studyObject{traccar}	&	193	&	22.4\%	\\
\studyObject{Commons Imag}	&	244	&	21.4\%	\\
\studyObject{openwayback}	&	166	&	18.6\%	\\
\studyObject{JFreechart}	&	754	&	17.7\%	\\
\studyObject{Google HTTP}	&	145	&	16.5\%	\\
\studyObject{javaparser}	&	293	&	14.0\%	\\
\studyObject{graphhopper}	&	252	&	11.5\%	\\
\studyObject{geometry API}	&	224	&	 9.9\%	\\
\studyObject{vectorz}	&	339	&	8.0\%	\\
\studyObject{jackson-db}	&	307	&	7.8\%	\\
\studyObject{bitcoinj}	&	77	&	4.7\%	\\
\studyObject{jsoup}	&	37	&	4.4\%	\\
\studyObject{urban-airship}	&	78	&	3.5\%	\\
\studyObject{Commons Geom}	&	20	&	2.8\%	\\
\studyObject{gson}	&	15	&	2.8\%	\\
\studyObject{Commons Math}	&	129	&	2.7\%	\\
\studyObject{Commons Stat}	&	7	&	2.6\%	\\
\studyObject{Commons Lang}	&	43	&	1.7\%	\\
\midrule
\textit{median}	&	\textit{166}	&	\textit{ 9.9\%}	\\

		\bottomrule
	\end{tabular}
	\label{Tbl:Mutation_Analysis_Infos}
\end{table}

Before addressing the research questions,
 we present in Table~\ref{Tbl:Mutation_Analysis_Infos} the absolute and relative number of \ineffMeth of each project
 as computed in the mutation analysis.
Depending on the project, between \minIneffTested and \maxIneffTested of the \textit{covered} methods are \ineffMeth.
According to these measurements, methods in \studyObject{gson} and four of the Apache projects are especially well tested compared to the other projects.
In contrast, the results of \studyObject{scifio}, \studyObject{pdfbox}, and \studyObject{biojava} are below average. 

\halfBaselinespace

\textbf{\rqNumSDist: \rqTextSDist}
Table~\ref{Tbl:SDist_Correlation_Results} shows the results of the \correlationMethod correlation test 
 between a method's \sdist and \mtRes.

\begin{table}
	\centering
	\caption{\rqNumSDist: \correlationMethod's correlation coefficient for a method's mutation result and its \sdist.
						\formatCaptionDetails{Absolute coefficient values $\geq 0.2$ and p-values $<0.05$ are highlighted.}}
	\begin{tabular}{lR{1.6cm}R{1.2cm}}
			\toprule
Project			& coefficient $ \downarrow $ & p-value \\
		\midrule
		
\studyObject{JFreechart}	&	\textbf{+0.58}	&	\cellcolor{ColorHighlightLightBlue} \textless 0.001 \\
\studyObject{scifio}	&	\textbf{+0.48}	&	\cellcolor{ColorHighlightLightBlue} \textless 0.001 \\
\studyObject{javaparser}	&	\textbf{+0.41}	&	\cellcolor{ColorHighlightLightBlue} \textless 0.001 \\
\studyObject{Commons Stat}	&	\textbf{+0.35}	&	\cellcolor{ColorHighlightLightBlue} \textless 0.001 \\
\studyObject{traccar}	&	\textbf{+0.33}	&	\cellcolor{ColorHighlightLightBlue} \textless 0.001 \\
\studyObject{pdfbox}	&	\textbf{+0.31}	&	\cellcolor{ColorHighlightLightBlue} \textless 0.001 \\
\studyObject{biojava}	&	\textbf{+0.29}	&	\cellcolor{ColorHighlightLightBlue} \textless 0.001 \\
\studyObject{graphhopper}	&	\textbf{+0.24}	&	\cellcolor{ColorHighlightLightBlue} \textless 0.001 \\
\studyObject{Commons Lang}	&	\textbf{+0.21}	&	\cellcolor{ColorHighlightLightBlue} \textless 0.001 \\
\studyObject{bitcoinj}	&	\textbf{+0.20}	&	\cellcolor{ColorHighlightLightBlue} \textless 0.001 \\
\studyObject{jackson-db}	&	+0.18	&	\cellcolor{ColorHighlightLightBlue} \textless 0.001 \\
\studyObject{jsoup}	&	+0.18	&	\cellcolor{ColorHighlightLightBlue} \textless 0.001 \\
\studyObject{Commons Geom}	&	+0.17	&	\cellcolor{ColorHighlightLightBlue} \textless 0.001 \\
\studyObject{Commons Imag}	&	+0.16	&	\cellcolor{ColorHighlightLightBlue} \textless 0.001 \\
\studyObject{geometry API}	&	+0.15	&	\cellcolor{ColorHighlightLightBlue} \textless 0.001 \\
\studyObject{openwayback}	&	+0.14	&	\cellcolor{ColorHighlightLightBlue} \textless 0.001 \\
\studyObject{gson}	&	+0.13	&	\cellcolor{ColorHighlightLightBlue} 0.003 \\
\studyObject{urban-airship}	&	+0.11	&	\cellcolor{ColorHighlightLightBlue} \textless 0.001 \\
\studyObject{Commons Math}	&	+0.08	&	\cellcolor{ColorHighlightLightBlue} \textless 0.001 \\
\studyObject{vectorz}	&	+0.07	&	\cellcolor{ColorHighlightLightBlue} \textless 0.001 \\
\studyObject{Google HTTP}	&	-0.17	&	\cellcolor{ColorHighlightLightBlue} \textless 0.001 \\

		\bottomrule
	\end{tabular}
	\label{Tbl:SDist_Correlation_Results}
\end{table}

We observe that a statistically significant correlation exists
 in all \numberOfStudyObjectsText projects (p-value $<0.05$).
The positive correlation coefficients indicate
 that the proportion of ineffectively tested methods increases with increasing \sdistShort values.
The strongest correlation is achieved in the project \studyObject{JFreechart} with a correlation coefficient of $\sdistMaxCorr$.
When looking at this project's test code, it was striking that the test cases contain many assertions.
A moderate correlation with a coefficient between 0.3 and 0.5 is present in five further projects.
A weak correlation is present in the remaining projects.
In the project \studyObject{Google HTTP} a weak negative correlation is observed;
 however, in this project, the \sdist does not exceed the value 2 in 81\% of the methods.

The red line in Figure~\ref{Fig:SDist_Diagram} presents the proportion of \ineffMeth per \sdist value.
In the project \studyObject{JFreechart}, more than 50\% of the methods with a \sdist higher than 3 are \ineff.
\definecolor{Color_Ineff_Proportion}{RGB}{255, 0, 0}
\definecolor{Color_Method_Execution_Proportion}{RGB}{217, 217, 217} 
\newcommand{\xAxisCropValue}{0.5\%\xspace}

\begin{figure*}
	\centering
	\includegraphics[width=1.0\linewidth]{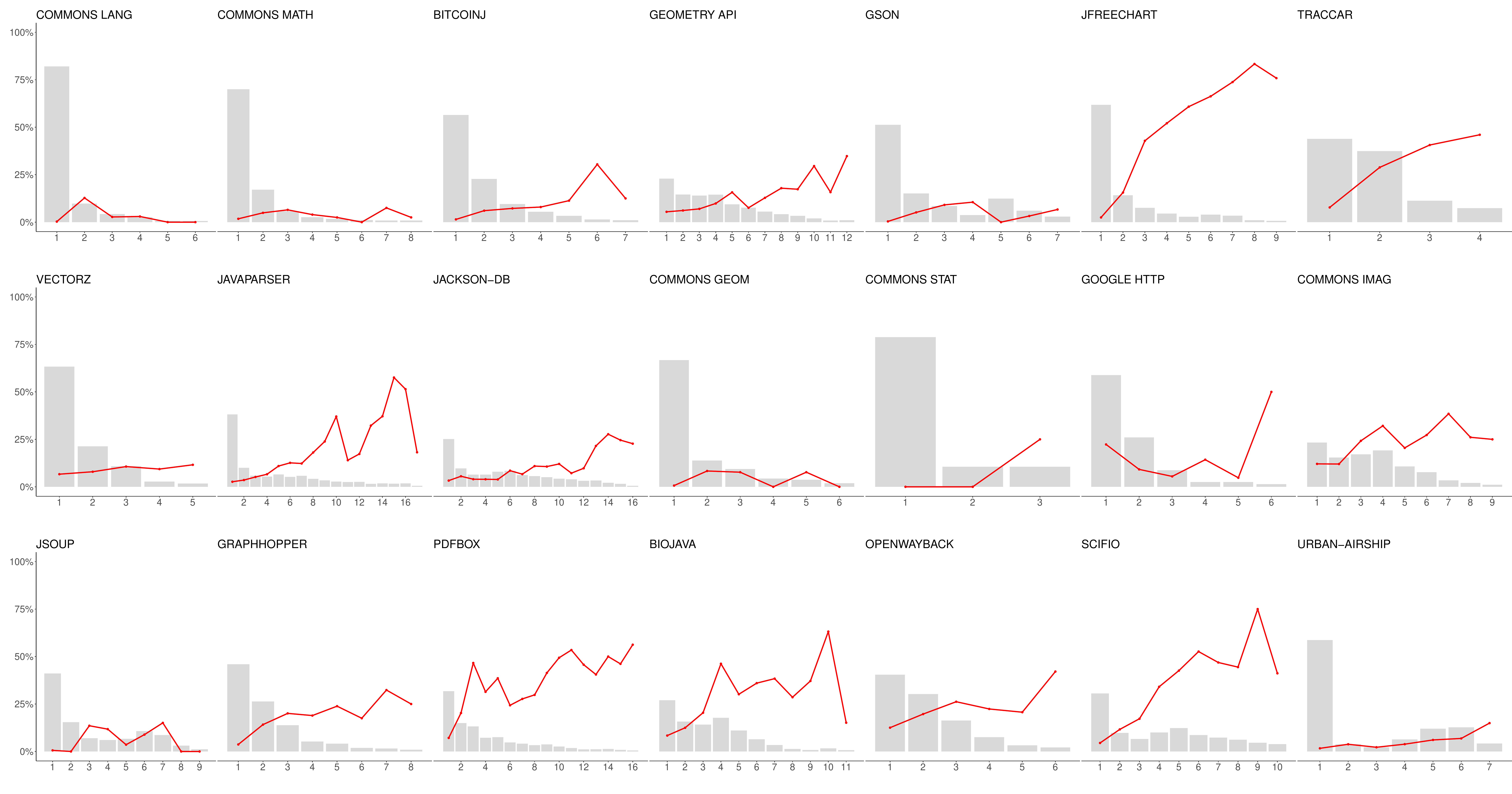}
	\caption[Proportion of \ineffMeth and of executed mutants per \sdist value.]{
						\rqNumSDist: The charts present
						the \coloredLine{Color_Ineff_Proportion}~proportion of \ineffMeth as red line
						and the \coloredSquare{Color_Method_Execution_Proportion}~proportion of methods per \sdist value as gray bars.
						\formatCaptionDetails{The hypothesis is that the proportion of ineffectively tested methods increases with increasing \sdist values.
						The x-axis is cropped when the proportion of methods per distance value falls below \xAxisCropValue.}
						}
	\label{Fig:SDist_Diagram}
\end{figure*}

\begin{figure}
	\centering
	\includegraphics[width=1.0\linewidth]{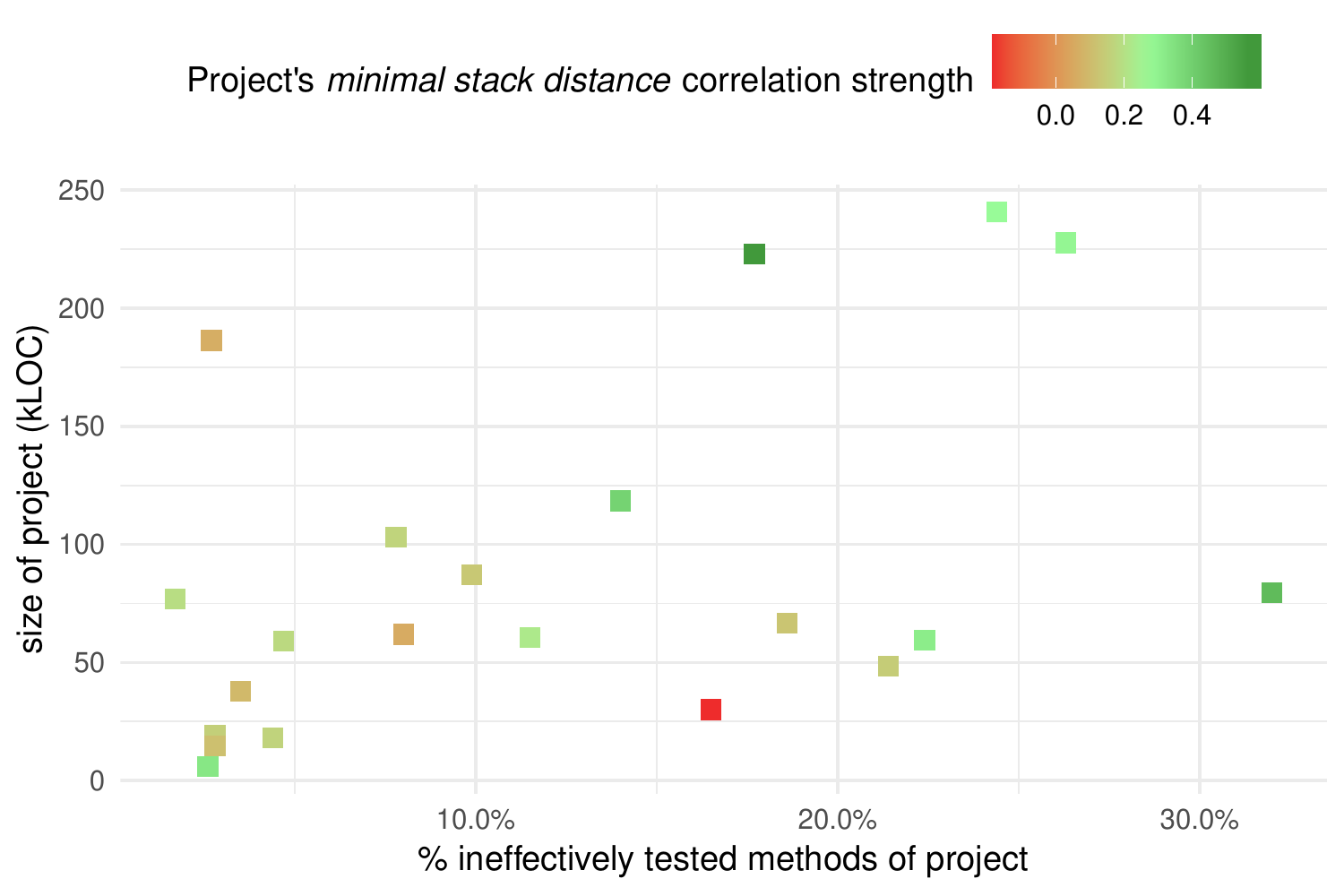}
	\caption{The projects' proportion of \ineffMeth (x-axis),
					 project size in \kloc (y-axis),
					 and the strength of the correlation between a method's \sdist and its \mtRes
					 from Table~\ref{Tbl:SDist_Correlation_Results} (color).}
	\label{Fig:SDist_Project_Correlation_Explanation}
\end{figure}

The illustration in Figure~\ref{Fig:SDist_Project_Correlation_Explanation} indicates that
 the correlation between a method's \sdist and its \mtRes 
 is stronger in larger projects with a high proportion of \ineffMeth.
(The correlation between the project's correlation coefficient and these two project characteristics is each 0.4.)

\formatResultBox{Methods with a higher \sdist to covering test cases are more likely to be \ineff.}

\halfBaselinespace

\textbf{\rqNumPrediction: \rqTextPrediction}

\begin{table}
	\centering
	\caption{\rqNumPrediction: Performance when predicting a \textit{method's mutation result.}}
	\begin{tabular}{lR{1.4cm}R{1.4cm}R{1.4cm}}
		\toprule
Project & Precision & Recall & F-score $ \downarrow $ \\
		\midrule

\studyObject{Commons Stat}	&	 99.9\%	&	 99.9\%	&	 99.9\%	\\
\studyObject{Commons Lang}	&	98.8\%	&	98.9\%	&	98.7\%	\\
\studyObject{gson}	&	97.5\%	&	97.7\%	&	97.1\%	\\
\studyObject{Commons Math}	&	96.7\%	&	97.5\%	&	96.7\%	\\
\studyObject{Commons Geom}	&	96.2\%	&	97.2\%	&	96.4\%	\\
\studyObject{urban-airship}	&	96.3\%	&	96.9\%	&	96.3\%	\\
\studyObject{Google HTTP}	&	95.1\%	&	95.1\%	&	94.9\%	\\
\studyObject{jsoup}	&	94.1\%	&	95.6\%	&	94.3\%	\\
\studyObject{bitcoinj}	&	93.7\%	&	95.3\%	&	94.0\%	\\
\studyObject{JFreechart}	&	93.1\%	&	93.4\%	&	93.1\%	\\
\studyObject{javaparser}	&	92.9\%	&	93.2\%	&	92.8\%	\\
\studyObject{vectorz}	&	92.5\%	&	93.5\%	&	92.4\%	\\
\studyObject{jackson-db}	&	91.5\%	&	93.0\%	&	91.7\%	\\
\studyObject{graphhopper}	&	89.5\%	&	90.8\%	&	89.3\%	\\
\studyObject{geometry API}	&	86.6\%	&	90.0\%	&	87.1\%	\\
\studyObject{traccar}	&	86.8\%	&	87.1\%	&	86.9\%	\\
\studyObject{Commons Imag}	&	87.2\%	&	87.7\%	&	86.8\%	\\
\studyObject{biojava}	&	85.1\%	&	85.7\%	&	85.1\%	\\
\studyObject{pdfbox}	&	84.1\%	&	84.7\%	&	83.8\%	\\
\studyObject{openwayback}	&	81.3\%	&	83.5\%	&	81.4\%	\\
\studyObject{scifio}	&	78.7\%	&	79.0\%	&	78.8\%	\\
\midrule
\textit{median}	&	\textit{92.9\%}	&	\textit{93.4\%}	&	\textit{92.8\%}	\\

	\bottomrule
	\end{tabular}
	\label{Tbl:Prediction_Performance_MXO_both_within}
\end{table}

Table~\ref{Tbl:Prediction_Performance_MXO_both_within} presents the classifier's precision, recall, and F-score
 of the within-project prediction of \predGoal.
As described in Section~\ref{Sec:Study_Design},
 the performance measures constitute the weighted average of the outcomes \textit{\ineffShort} and \textit{\effectT}.
Median precision is \predictionWithinMxoForBothMedianPrecision, and median recall is \predictionWithinMxoForBothMedianRecall.
When conducting cross-project prediction for the same scenario,
 median precision and recall deteriorate to \predictionCrossMxoForBothMedianPrecision resp. \predictionCrossMxoForBothMedianRecall.

\begin{table}
	\centering
	\caption{\rqNumPrediction: Performance when predicting \textit{\ineffMeth.}}
	\begin{tabular}{lR{1.4cm}R{1.4cm}R{1.4cm}}
		\toprule
Project & Precision & Recall & F-score $ \downarrow $ \\
		\midrule

\studyObject{Commons Stat}	&	96.6\%	&	100.0\%	&	98.2\%	\\
\studyObject{Google HTTP}	&	94.6\%	&	74.8\%	&	83.5\%	\\
\studyObject{JFreechart}	&	87.0\%	&	73.4\%	&	79.6\%	\\
\studyObject{javaparser}	&	84.1\%	&	63.4\%	&	72.3\%	\\
\studyObject{traccar}	&	72.6\%	&	68.0\%	&	70.2\%	\\
\studyObject{biojava}	&	76.4\%	&	59.9\%	&	67.1\%	\\
\studyObject{pdfbox}	&	78.4\%	&	57.6\%	&	66.4\%	\\
\studyObject{scifio}	&	68.5\%	&	63.8\%	&	66.1\%	\\
\studyObject{Commons Imag}	&	81.1\%	&	55.5\%	&	65.9\%	\\
\studyObject{Commons Lang}	&	85.5\%	&	41.3\%	&	55.7\%	\\
\studyObject{graphhopper}	&	70.6\%	&	34.3\%	&	46.2\%	\\
\studyObject{vectorz}	&	70.7\%	&	32.5\%	&	44.5\%	\\
\studyObject{openwayback}	&	60.7\%	&	32.5\%	&	42.4\%	\\
\studyObject{urban-airship}	&	64.2\%	&	27.6\%	&	38.6\%	\\
\studyObject{jackson-db}	&	61.4\%	&	26.6\%	&	37.1\%	\\
\studyObject{gson}	&	87.5\%	&	23.3\%	&	36.8\%	\\
\studyObject{Commons Math}	&	60.3\%	&	15.9\%	&	25.2\%	\\
\studyObject{Commons Geom}	&	50.0\%	&	15.0\%	&	23.1\%	\\
\studyObject{bitcoinj}	&	50.0\%	&	13.0\%	&	20.6\%	\\
\studyObject{jsoup}	&	51.5\%	&	11.5\%	&	18.8\%	\\
\studyObject{geometry API}	&	46.9\%	&	11.0\%	&	17.9\%	\\
\midrule
\textit{median}	&	\textit{70.7\%}	&	\textit{34.3\%}	&	\textit{46.2\%}	\\

	\bottomrule
	\end{tabular}
	\label{Tbl:Prediction_Performance_MXO_ineff_within}
\end{table}

Ineffectively tested methods represent the minority class and are therefore more difficult to predict.
Table~\ref{Tbl:Prediction_Performance_MXO_ineff_within} shows the within-project prediction performance
 for identifying \ineffMeth.
Median precision of this outcome is \predictionWithinMxoForIneffMedianPrecision and median recall is \predictionWithinMxoForIneffMedianRecall.
In the best case, \predictionWithinMxoForIneffBestCasePrecision precision and \predictionWithinMxoForIneffBestCaseRecall recall are still achieved (\studyObject{Commons Stat}).

\begin{figure}
	\centering
	\includegraphics[width=1.0\linewidth]{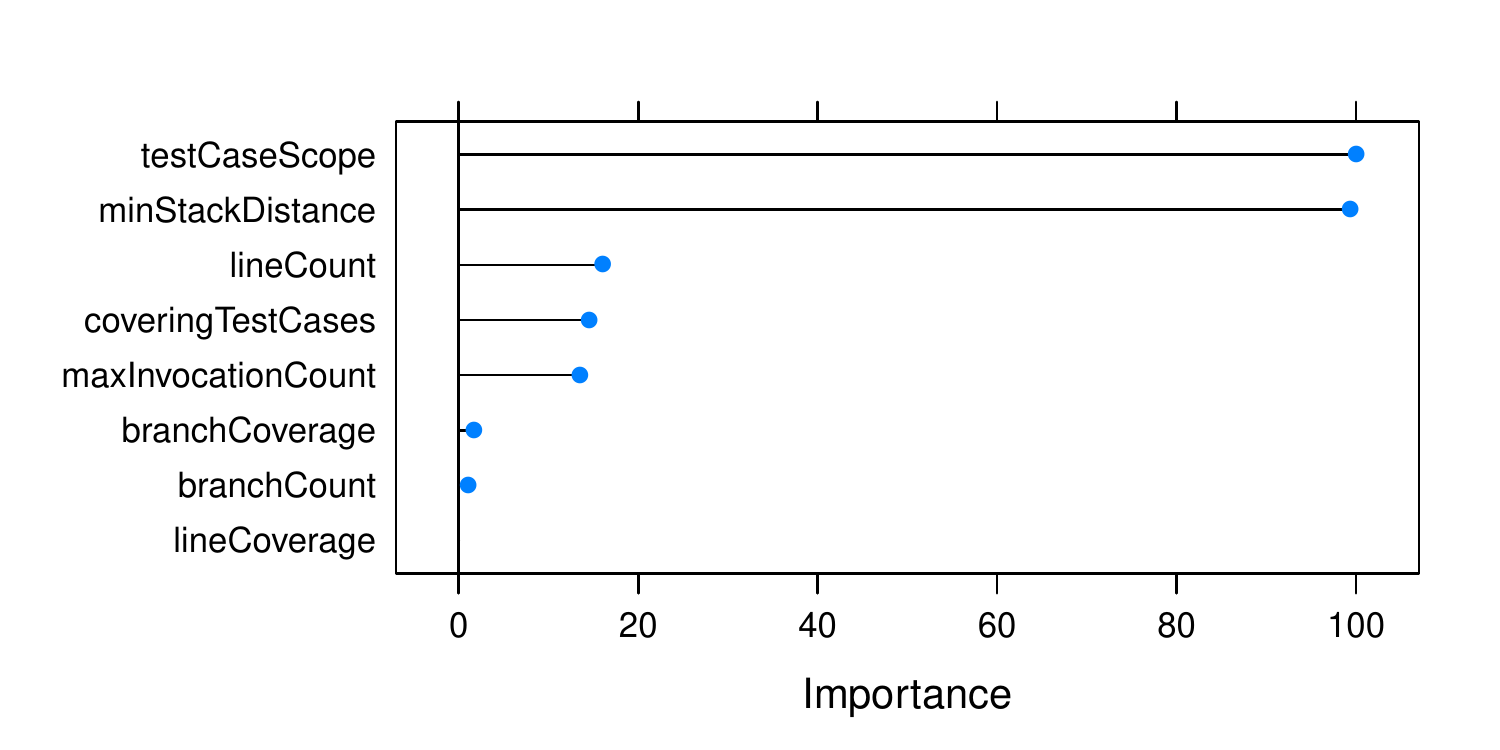}
	\caption{\rqNumPrediction: Variable importance of \studyObject{JFreechart}'s prediction model (scaled to one).}
	\label{Fig:Prediction_VarImp}
\end{figure}

Figure~\ref{Fig:Prediction_VarImp} exemplary presents the variable importance
 of \studyObject{JFree\-chart}'s within-project prediction model.
The figure shows that the \sdist
 and the minimal scope value of a method's covering test cases
 (the scope of a test case expresses how many methods it covers)
 are the most important variables for the prediction model.

Cross-project prediction for identifying \ineffMeth only achieves a poor performance.
Even when applying the over-sampling technique \textit{SMOTE}\footnote{Synthetic Minority Over-Sampling Technique~\cite{chawla2002smote}} to pre-process training sets,
 median precision is only \predictionCrossMxoForIneffWithSmoteMedianPrecision and median recall is \predictionCrossMxoForIneffWithSmoteMedianRecall.
Hence, cross-project prediction is not well suited for uncovering \ineffMeth.

\formatResultBox{
 The \mtRes of a method can on average be predicted
  with \predictionWithinMxoForBothMedianPrecision precision and \predictionWithinMxoForBothMedianRecall recall.
 Cross-project prediction is more challenging and achieves a weaker performance.
}

\begin{table}
	\centering
	\caption{\rqNumPrediction: Performance when predicting the mutation result \textit{of a method test-case pair.}}
	\begin{tabular}{lR{1.4cm}R{1.4cm}R{1.4cm}}
		\toprule
Project & Precision & Recall & F-score $ \downarrow $ \\
		\midrule
\studyObject{scifio}	&	92.8\%	&	92.8\%	&	92.8\%	\\
\studyObject{Commons Stat}	&	92.1\%	&	92.4\%	&	91.7\%	\\
\studyObject{Commons Geom}	&	90.8\%	&	91.2\%	&	90.4\%	\\
\studyObject{javaparser}	&	90.1\%	&	90.2\%	&	90.1\%	\\
\studyObject{urban-airship}	&	89.1\%	&	90.2\%	&	88.7\%	\\
\studyObject{Google HTTP}	&	88.4\%	&	88.6\%	&	88.2\%	\\
\studyObject{gson}	&	87.9\%	&	88.0\%	&	87.9\%	\\
\studyObject{Commons Lang}	&	87.5\%	&	87.8\%	&	86.8\%	\\
\studyObject{JFreechart}	&	86.5\%	&	86.5\%	&	86.4\%	\\
\studyObject{bitcoinj}	&	86.1\%	&	86.1\%	&	86.1\%	\\
\studyObject{Commons Math}	&	85.2\%	&	85.7\%	&	85.1\%	\\
\studyObject{traccar}	&	85.1\%	&	85.0\%	&	85.1\%	\\
\studyObject{vectorz}	&	85.4\%	&	86.6\%	&	84.9\%	\\
\studyObject{jsoup}	&	84.4\%	&	84.9\%	&	84.1\%	\\
\studyObject{pdfbox}	&	83.8\%	&	83.8\%	&	83.8\%	\\
\studyObject{Commons Imag}	&	82.5\%	&	82.7\%	&	82.2\%	\\
\studyObject{biojava}	&	81.7\%	&	81.7\%	&	81.5\%	\\
\studyObject{openwayback}	&	80.9\%	&	80.8\%	&	80.8\%	\\
\studyObject{graphhopper}	&	80.6\%	&	80.7\%	&	80.5\%	\\
\studyObject{geometry API}	&	77.6\%	&	78.1\%	&	77.4\%	\\
\studyObject{jackson-db}	&	72.4\%	&	72.4\%	&	72.4\%	\\
\midrule
\textit{median}	&	\textit{85.4\%}	&	\textit{86.1\%}	&	\textit{85.1\%}	\\

	\bottomrule
	\end{tabular}
	\label{Tbl:Prediction_Performance_MTO_both_within}
\end{table}

The above results concern the prediction of a method's \mtRes with respect to all test cases.
For other use cases, e.g., for enhancing test case prioritization with test effectiveness information,
 it can also be useful to predict the \mtRes of a method test-case \textit{pair.}
Table~\ref{Tbl:Prediction_Performance_MTO_both_within} presents the within-project performance
 when predicting the \mtRes of a method test-case pair.
In this scenario, median precision and recall are \predictionWithinMtoPairForBothMedianPrecision resp. \predictionWithinMtoPairForBothMedianRecall.
When focusing on the outcome \textit{\ineff,} median precision and recall still achieve \predictionWithinMtoPairForIneffMedianPrecision resp. \predictionWithinMtoPairForIneffMedianRecall.

Hence, the prediction achieves promising results when working on method test-case pairs.
A reason for this is that, unlike when predicting the result of a method with respect to all test cases,
 test case metrics are not aggregated.

\formatResultBox{
 Ineffectively tested method test-case pairs can be predicted
  with \predictionWithinMtoPairForIneffMedianPrecision precision and \predictionWithinMtoPairForIneffMedianRecall recall on average.
}

Zhang et al.~\cite{zhang2018predictive} achieved precision and recall values of around 90\%
 (depending on project and scenario).
They only present performance measures aggregated of both outcomes.
Although an in-depth comparison with their results does not seem sensible---because
 they predicted for different mutation operators,
 used other metrics,
 and included methods not covered by any test---we
 can still say that the prediction performance is roughly comparable.

\section{Discussion}
\label{Sec:Discussion}

The study's results show that the correlation between a method's \sdist and its \mtRes 
 is moderate to strong in six projects and present in further projects to a lower degree.
In general, the correlation is stronger in larger projects
 (\studyObject{JFreechart}, \studyObject{biojava}, \studyObject{pdfbox}),
 which also exhibit higher \sdist values.
In large, multi-module projects some methods are only tested by integration tests,
 which usually have a higher distance to many of the covered methods than a unit test does.
In such projects, the \sdist can provide valuable insights about the testing state of methods
 and thereby provide an additional value to coverage information.

The evaluation of the prediction models shows that
 machine learning models can successfully predict the \mtRes of a method.
Hence, such models can be considered as a light-weight alternative to mutation testing.
\begin{table}
	\centering
	\caption{Duration of analyses (in hours) and slowdown factor based on the normal test suite execution.}
	\resizebox{\linewidth}{!}{
  \begin{tabular}{lR{1.4cm}R{1.5cm}R{1.35cm}R{1.35cm}}
	
\toprule
Project       & Test Suite Execution         & Test Suite Execution + Stack Dist. Recording & Mutation Analysis with Early Abort & Mutation Analysis with Full Matrix \\
\midrule

 \studyObject{biojava}				& 00:27:45				& 01:31:00				& 23:00:00 					& 46:49:00 \\
															& \textit{(1.0)} 	& \textit{(3.3)} 	& \textit{(49.7)} 	& \textit{(101.2)} \\
 \studyObject{bitcoinj}				& 00:01:40				& 00:02:45				& 00:43:26 					& 03:36:00 \\
															& \textit{(1.0)} 	& \textit{(1.7)} 	& \textit{(26.1)} 	& \textit{(129.6)} \\
 \studyObject{JFreechart}			& 00:00:13 				&	00:00:17 				& 00:09:07					& 00:13:38 \\
															& \textit{(1.0)} 	& \textit{(1.3)} 	& \textit{(42.1)} 	& \textit{(63.0)} \\
 \studyObject{pdfbox}					&	00:01:38				& 00:07:56				&	02:33:00					& 05:14:00 \\
															& \textit{(1.0)} 	& \textit{(4.9)} 	& \textit{(93.7)} 	& \textit{(192.0)} \\

\bottomrule

\end{tabular}
}
	\label{Tbl:Analysis_Duration}
\end{table}

To point out possible time savings,
 Table~\ref{Tbl:Analysis_Duration} presents the duration of different analyses exemplarily of four projects.
The current---not yet performance-optimized---implementation for recording the \sdist has an influence on the duration of the test execution.
 It slows the execution down by a low but perceptible single-digit factor.
Nonetheless, a prediction model using this metric can achieve significant savings compared to the execution of a mutation analysis.
The analysis with the state-of-the-art mutation testing tool \pit takes about 50--200 times as long as a single execution of the corresponding test suite.
In the largest project (\studyObject{biojava}),
 the computation of a full mutation matrix took more than 46 hours (101 times the duration of the test execution) and
 an analysis that stops assessing a mutant after having found the first killing test case still needed 23 hours.
Consequently, such prediction models can also be taken into consideration in projects
 in which a mutation analysis is not applicable due to a long duration.

\section{Threats to Validity}
\label{Sec:Threats_To_Validity}

We separate the threat to validity into internal and external threats.

The computation of the stack distance is a threat to internal validity.
Although we developed the computation logic with great care,
 the implementation could contain faults that affect the outcome.
To mitigate this threat, we verified computed values of different code samples
 and developed automated tests to check the implementation.
In addition, the source code of our tool can be inspected on GitHub~\citeTool.

The same applies to the conducted extension of the \pit mutation testing tool
 to enable computing a full mutation matrix.
To mitigate this threat, we created a pull request, which was carefully reviewed and merged by the head developer of \pit~\cite{Niedermayr2018MutationMatrixPR}.

Some of the generated mutants may be equivalent mutants,
 which differ only syntactically but not semantically from the original source code,
 and, hence, cannot be killed~\cite{grun2009impact}.
Therefore, some of the mutants that were regarded as surviving
 could be equivalent mutants and affect the results.
Due to the design of the mutation operator (cf. Section~\ref{Sec:Descartes_Operator})
 and the exclusion of empty methods and methods returning \codeText{null},
 hardly any equivalent mutants are generated~\cite{niedermayr2016teststellme}.
A manual review on a sample confirmed this observation.

Although we selected \numberOfStudyObjectsText study objects with different characteristics,
 the selection of the projects poses a threat to external validity.
Since we chose only open-source projects that use Maven as build system and in which nearly all tests succeed,
 well-engineered projects with mature test suites may be over-represented in our sample.
Hence, future work is necessary to validate whether the results are generalizable for Java projects
 and projects in other programming languages.

\section{Conclusion}
\label{Sec:Conclusion}

In this paper, we proposed and studied the \sdist measure,
 which describes the proximity of a method to any of its test cases.
Our results indicate that a correlation exists between this measure
 and a property indicating whether a method is \ineff (pseudo-tested).
Classifiers that predict \predGoal
 achieve a median precision of \predictionWithinMxoForBothMedianPrecision and recall of \predictionWithinMxoForBothMedianRecall.
The measures needed for such a classifier can be computed in a single test suite execution,
 while mutation testing may take---depending on the size of an application---several hours or days.
Therefore, we suggest considering such classifiers
 as \usageLong.
In particular, the classifiers can be a reasonable alternative in continuous integration.
Furthermore, they can be useful for projects in which a mutation analysis is not applicable
 (due to the analysis duration or class loading issues).

For future work, 
 we plan to investigate more measures, such as, information about assertions in tests,
 and incorporate them into the prediction models to further improve their performance.
In addition, we want to enhance cross-project predictions.
For that, we plan to include project characteristics into the model
 and focus model training on projects with similar properties.

\begin{acks}
	\acknowledgmentText
\end{acks}

\balance

\bibliographystyle{ACM-Reference-Format}
\bibliography{./literature/bibtex}

\end{document}